\documentclass[journal]{IEEEtran}
\ifCLASSINFOpdf
   \usepackage[pdftex]{graphicx}

\else

\fi

\usepackage{xcolor}
\usepackage{layouts}
\usepackage{mathtools}
\usepackage{cuted}
\usepackage{booktabs}
\usepackage{amsfonts}
\usepackage{amssymb}
\usepackage{subfig}
\usepackage{amsmath}
\usepackage{cite}
\usepackage{romannum}
\usepackage{amssymb, amsmath, amsthm, amsfonts}
\newcommand{\btheta}{\boldsymbol{\theta}}
\newcommand{\bmu}{\boldsymbol{\mu}}
\newcommand{\bV}{\mathbf{V}}

\newcommand{\init}{\text{init}}

\newcommand{\st}{\text{s.t.}}

\begin{document}
\title{An Iterative Approach to Finding Global Solutions of AC Optimal Power Flow Problems
}

\author{Ling Zhang,~\IEEEmembership{Student Member,~IEEE,}
        and~Baosen~Zhang,~\IEEEmembership{Member,~IEEE,}%
\thanks{L. Zhang and B. Zhang are with the Department
of Electrical and Computer Engineering, University of Washington, WA, 98195 USA e-mail: \{lzhang18,zhangbao\}@uw.edu 
The authors are partially supported by NSF grants ECCS-1942326 and ECCS-2023531}
}

\maketitle

\begin{abstract}
The existence of multiple solutions to AC optimal power flow (ACOPF) problems has been noted for decades. Existing solvers are generally successful in finding local solutions, which are stationary points but may not be globally optimal. In this paper, we propose a simple iterative approach to find globally optimal solutions to ACOPF problems. 
First, we call an existing solver for the ACOPF problem. From the solution and the associated dual variables, we form a partial Lagrangian. Then we optimize this partial Lagrangian and use its solution as a warm start to call the solver again for the ACOPF problem. 
By repeating this process, we can iteratively improve the solution quality, moving from local solutions to global ones.
We show the effectiveness our algorithm on standard IEEE networks. The simulation results show that our algorithm can escape from local solutions to achieve global optimums within a few iterations.

\end{abstract}


\IEEEpeerreviewmaketitle

\section{Introduction}

Optimal power flow (OPF) is a fundamental resource allocation problem in power system operations that minimizes the cost of power generation while satisfying demand. The ACOPF formulation of the problem uses nonlinear power flow equations, resulting in nonlinear and nonconvex optimization problems~\cite{cain2012history,Molzahn19,hiskens2001exploring}.  
The consequence of the nonconvexity of ACOPF we study in this paper is the presence of multiple solutions. The existence of multiple local solutions of the OPF problem has been well-known for several decades~\cite{momoh1999review}. A common assumption is that OPF problems tend to have a single ``practical'' solution, and therefore the fact that multiple solutions can exist do not impact day-to-day operations~\cite{momoh1997challenges,Wei98}. However, an increasingly large body of work have pointed to that multiple solutions do occur under reasonable conditions and cannot easily ruled out~\cite{bukhsh2013local,wu2017deterministic,Molzahn19}. For example, \cite{bukhsh2013local} shows how modifications of the standard IEEE benchmarks can lead to each having more than one local solutions.

Most ACOPF problems are solved via variations of nonlinear optimization algorithms, including Newton-Raphson, sequential programming, interior points and others (see~\cite{Molzahn19,qiu2009literature,capitanescu2016critical} and the references within). These algorithms are in general only able to certify a solution is locally optimal. These solutions satisfy some necessary optimality conditions (e.g., the KKT conditions) but may not globally optimal, that is, the least cost solution. An open question in the field is to develop algorithms that can find global optimal solutions. In addition, understanding and distinguishing the local and the global optimal solutions can lead to important theoretical discoveries about the ACOPF problem.

\begin{figure}[ht]
    \centering
    \includegraphics[scale=0.4]{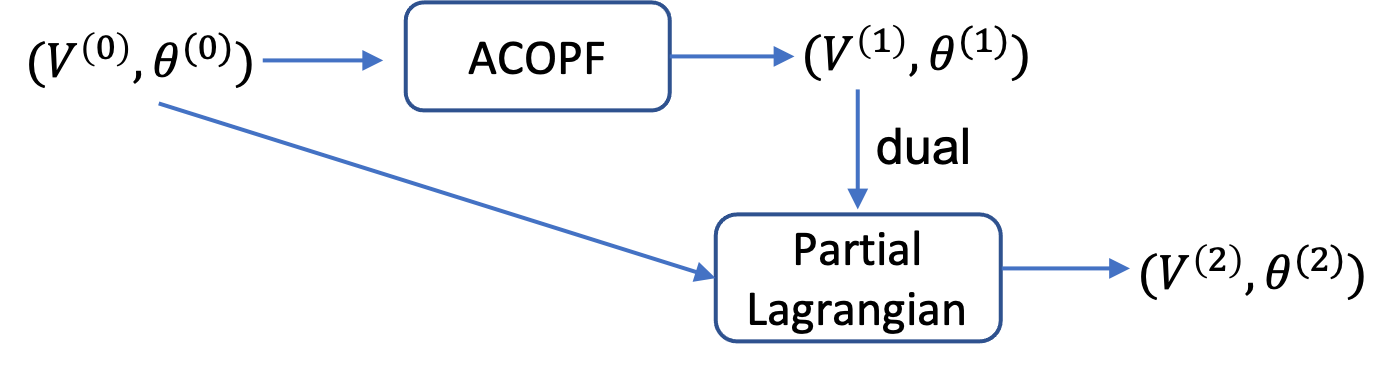}
    \caption{Outline of the solution process.}
    \label{fig:intro}
\end{figure}

In this paper, we propose a simple algorithm that can effectively escape from local solutions to seek global optimal solutions. The process is outlined in Fig.~\ref{fig:intro}. First, we solve the ACOPF problem using an existing solver (e.g., IPOPT~\cite{wachter2006implementation}). From the solution and its associated dual variables, we form a partial Lagrangian. This partial Lagrangian serves to ``flatten'' the geometry of the optimization problem. We then optimize this partial Lagrangian, which can lead to a different solution. Using this second solution as a warm start, we again call the solver for the ACOPF problem. Repeating this iterative process, we can successively improve the solution quality, moving from local solutions to globally optimal ones. 

We show the performance of our algorithm on standard 9-bus, 22-bus and 39-bus networks.
We show that our algorithm can escape from local solutions~\cite{bukhsh2013local} and find the global optimal solution a single iteration for most instances. 
We also sketch the intuition behind the algorithm using a two bus system.

Our approach can be seen as a way to provide good warm starts to nonlinear optimization solvers. Existing approaches along this line either randomizes~\cite{CE21} or uses a previous solution as the starting point~\cite{tang2017distributed}. The former tends to be time consuming, while the latter does not guarantee the resulting solution would be globally optimal~\cite{bukhsh2013local}. The work in \cite{mulvaney2020load} suggests that solutions can escape local minima if load undergoes random fluctuations. Our approach can be seen as providing an explicit and deterministic algorithm to search for global solutions by using dual variables at local solutions. 


\section{Model and Problem Formulation} \label{sec:model}
Consider a power system network where $n$ buses are connected by $m$ edges. 
For bus $i$, let $P^{G}_i$ and $Q^{G}_i$ denote the active and reactive output of the generated and $P^{D}_i$ and $Q^{D}_i$ denote the active and reactive load. We use $P^{f}_{ij}$ and $Q^{f}_{ij}$ to denote the active and reactive power flowing from bus $i$ to bus $j$. Note that if buses $i$ and $j$ are not connected, then the flows are zero.
We use $\theta_{ij}$ as a shorthand for $\theta_i-\theta_j$. 

The OPF problem is to minimize the total of active power generation costs while satisfying a number of constraints:
\begin{subequations} 
\label{prob1}
\begin{align}
     \min_{\bV, \btheta} &\textstyle \sum_{i} c_i(P^{G}_i)\\
    \st ~ & P^{G}_i = P^{D}_i + \textstyle \sum_{j=1}^{N} P^{f}_{ij}\label{Pbalanc}\\
    & Q^{G}_i = Q^{D}_i + \textstyle \sum_{j=1}^{N} Q^{f}_{ij}\label{Qbalanc}\\
    & P^{f}_{ij} = V_i^2g_{ij}-V_iV_j(g_{ij}\cos(\theta_{ij})-b_{ij}\sin(\theta_{ij}))\label{PEq}\\
    & Q^{f}_{ij} = V_i^2 \hat{b}_{ij} - V_iV_j(b_{ij}\cos(\theta_{ij}) +g_{ij}\sin(\theta_{ij}))\label{QEq}\\
    & \underbar{V}_i\leq V_i \leq \bar{V}_i\label{Vlimits}\\
    & \underbar{P}^{G}_i\leq P^G_i \leq \bar{P}^{G}_i\label{PGlimits}\\
    & \underbar{Q}^{G}_i\leq Q^G_i \leq\bar{Q}^{G}_i\label{QGlimits}\\
    & (P^{f}_{ij})^2+(Q^{f}_{ij})^2\leq (S_{ij}^{\max})^2\label{flimits}
\end{align}
\end{subequations}
where $\hat{b}_{ij}=b_{ij}-0.5b_{ij}^C$ and $b_{ij}^C$ is the line charging susceptance. The constraints  (\ref{Pbalanc}) and (\ref{Qbalanc}) enforce power balance, (\ref{PEq}) and (\ref{QEq}) are the AC power flow equations, (\ref{Vlimits}) limits the bus voltage magnitudes, (\ref{PGlimits}) and (\ref{QGlimits}) represent the active and reactive limits and \eqref{flimits} are the line flow limits. The cost at bus $i$ is $c_i(\cdot)$ and can be linear, quadratic or other functions. 


A pair of $(\bV, \btheta)$ that minimizes the objective cost among all feasible solutions to \eqref{prob1} is called a global solution. In practice,
a nonlinear programming (NLP) solver may only return a local solution, which is a feasible point that satisfy some local optimality conditions (e.g., KKT). Since a local solution is not necessarily global, in the following subsection, we propose an iterative approach to find global solutions by alternatively solving (\ref{prob1}) and its partial Lagrangian. Any existing NLP can be used, and and we use IPOPT in this paper.

\section{Algorithm} \label{sec:algo}
Our algorithm starts with a call to a NLP solver with an initial guess, denoted by $\btheta_{\init}$, $\bV_{\init}$. For example, this can be the standard flat start. 
Then we assume the solver returns a feasible solution. At this solution, we record the dual variables associated with the power balance equations \eqref{Pbalanc} and \eqref{Qbalanc}, denoted as $\bar{\bmu}^P$ and $\bar{\bmu}^Q$. Using these dual variables, we form the following partial Lagrangian by dualizing the power balance equations: 
\begin{subequations}
\begin{align}
\mathcal{L}(\bV, \btheta, \bmu^P, \bmu^Q)=& \textstyle \sum_{i} c_iP^{G}_i+ \textstyle \sum_{i}\mu_i^{P}(P^{D}_i + \textstyle \sum_{j=1}^{N} P^{f}_{ij} - P^{G}_i)\nonumber\\
    & + \textstyle \sum_{i}\mu_i^Q(Q^{D}_i + \sum_{j=1}^{N} Q^{f}_{ij} - Q^{G}_i). \nonumber 
\end{align}
\end{subequations}
We then minimize the partial Lagrangian by solving 
\begin{align} 
  \min_{\bV, \btheta} \; & \mathcal{L}(\bV, \btheta, \bmu^P, \bmu^Q) \label{prob2} \\
    \st & ~ (\ref{PEq})- (\ref{flimits}). \nonumber
\end{align}

We solve the problem in \eqref{prob2}
\emph{at the same initial point $(\bV_{\init}, \btheta_{\init})$} that was used to solve the original primal problem in \eqref{prob1}. Denote this solution to \eqref{prob2} by $(\bar{\bV}, \bar{\btheta})$. Then we start the solver again to solve (\ref{prob1}) but with the initial point $(\bar{\bV}, \bar{\btheta})$. This process can be repeated until the solutions stop changing or up to a predefined number of iterations.

It tures out that the initial point $(\bar{\bV}, \bar{\btheta})$ that is given by solving the partial Lagrangian is often a much better starting point than the original choice of $(\bV_{\init}, \btheta_{\init})$. Therefore, by repeating these steps, we can iteratively improve the quality of the starting point and get the global solution. The algorithm is summarized below as Algorithm 1. We illustrate the intuition behind this algorithm in the next section using a two bus network and present the numerical results in Section~\ref{sec:results}. 
\begin{table}[ht]
\normalsize
\begin{tabular}{ll}
\hline
\multicolumn{2}{l}{\textbf{Algorithm 1:  Solving ACOPF iteratively}}\\
\hline
\multicolumn{2}{l}{\textbf{Inputs:}~$\btheta_{\init}^{(0)}$, $\bV_{\init}^{(0)}$}\\
1:~At $i$-th iteration: Initialized at $\btheta_{\init}^{(i)}$, $\bV_{\init}^{(i)}$:\\
2:~Call IPOPT solver to solve problem \eqref{prob1}, record $(\bar{\bmu}^P_{(i)}, \bar{\bmu}^Q_{(i)})$\\
3:~Given $(\bar{\bmu}^P_{(i)}, \bar{\bmu}^Q_{(i)})$, call IPOPT for the partial Lagrangian in \eqref{prob2}, \\
~~~record the solutions as $(\bar{\btheta}^{(i)}, \bar{\bV}^{(i)})$\\
4:~Call IPOPT for \eqref{prob1} initialized at $(\bar{\btheta}^{(i)}$, $\bar{\bV}^{(i)})$ \\
~~~record solutions $(\hat{\btheta}^{(i)}, \hat{\bV}^{(i)})$\\
5:~If the solution from line 4 corresponds to lower \\
objective function value, then update initial points:\\
~~~$\btheta_{\init}^{(i+1)} = \hat{\btheta}^{(i)}$, $\bV_{\init}^{(i+1)} = \hat{\bV}^{(i)}$\\
6:~Otherwise, stay with the original initial points:\\
7:~Repeat the above process until solutions stop changing \\
~~~or reach the maximum number of iterations.\\
\hline
\end{tabular}
\label{algo}
\end{table}



\vspace{-0.5cm}
\section{Geometry}
\begin{figure}[t]
\centering
\includegraphics[height=5cm, width=6cm]{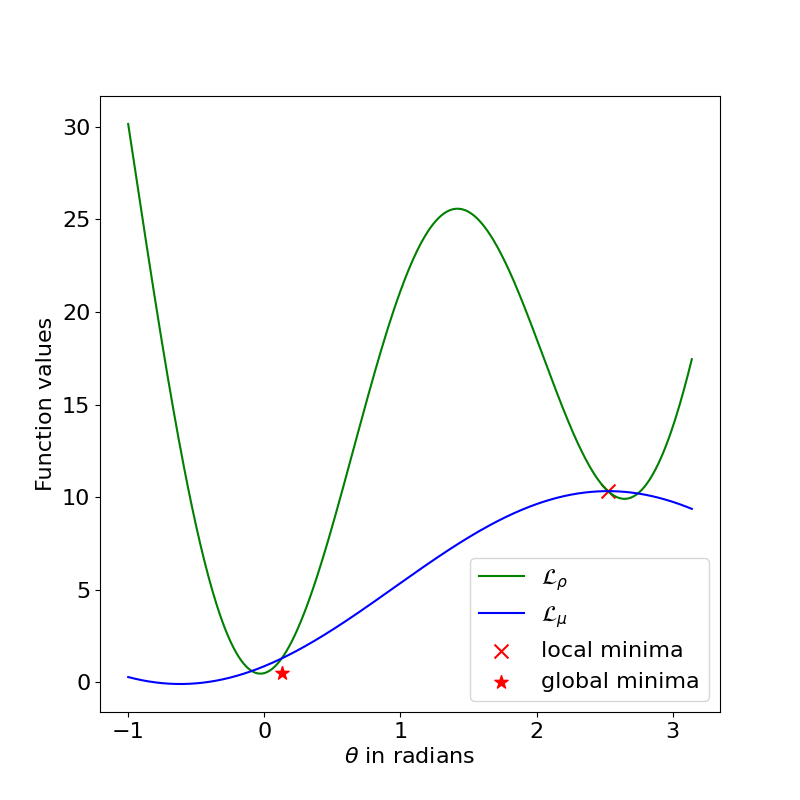}
\caption{Geometry of the penalized objective functions $\mathcal{L}_{\rho}$ and the partial Lagrangian $\mathcal{L}_{\mu}$.
\vspace{-0.5cm}}
\label{fig:2bus}
\end{figure}

Algorithm~1 is successful because of the geometry of the partial Lagrangian is ``better'' than the geometry of the original problem. We use a two-bus network as an illustrative example. 
For simplicity, we set both voltage magnitudes to 1 p.u. and optimize over the angle. Suppose bus 1 is a generator and the reference (slack) bus with linear cost at \$1/MW, and bus 2 is the load bus with angle $-\theta$. The line admittance is $g-jb$. Given a load of $l$ at bus 2, the ACOPF is:
\begin{subequations} \label{eqn:two_bus}
\begin{align}
    \min_{\theta} \; & g-g\cos(\theta)+b\sin(\theta)\\
    \st~& l+g-g\cos(\theta)-b\sin(\theta)=0\label{eq:2bus}. 
\end{align}
\end{subequations}
For a feasible load $l$, there are two solutions to \eqref{eq:2bus}. To make visualize how an nonlinear solver would approach \eqref{eqn:two_bus}, we convert it to a penalized unconstrained problem~\cite{bertsekas1997nonlinear}: 
\begin{align}
    \mathcal{L}_{\rho}= & g-g\cos(\theta)+b\sin(\theta) \label{eqn:penl}\\ 
    &+\rho/2(l+g-g\cos(\theta)-b\sin(\theta))^2, \nonumber
\end{align}
where $\rho$ is a (large) penalty parameter. The function $\mathcal{L}_{\rho}$ is plotted in Fig.~\ref{fig:2bus} and there are two local minima, with the left one being global. If a NLP is initialized in the right valley, the suboptimal solution would be found. 

Now consider the partial Lagrangian for \eqref{eqn:two_bus}, given by 
\begin{align} 
    \mathcal{L}_{\mu}= & g-g\cos(\theta)+b\sin(\theta) \label{eqn:Lag} \\
    & +\mu(l+g-g\cos(\theta)-b\sin(\theta)), \nonumber
\end{align}
where $\mu$ is the multiplier corresponding to the equality constraint (\ref{eq:2bus}) at a local solution. The blue line in Fig.~\ref{fig:2bus} plots $\mathcal{L}_{\mu}$ when dual variable is computed at the right (suboptimal) local solution. In contrast to $\mathcal{L}_{\rho}$, $\mathcal{L}_{\mu}$ has a unique minimum that falls into the \emph{left} valley, and can be obtained by minimizing $\mathcal{L}_{\mu}$ from any initialization. Therefore, if we use the minimum of $\mathcal{L}_{\mu}$ as an initialization point for the nonlinear solver, we would reach the global optimal solution. 

The above example illustrates that the Lagrangian has a more advantageous geometry for optimization than the original primal problems. This is not surprising, since equality constraints tend to complicate the optimization landscape~\cite{Molzahn19}. The key insight is to observe that it suffices to use the dual variables at \emph{suboptimal} local solutions to form the Lagrangian. 

\captionsetup[figure]{font=small,skip=2pt}




\vspace{-0.2cm}
\section{Simulation Results} \label{sec:results}
In this section we report the simulation results to validate the effectiveness of our algorithm.
The NLP solver used is IPOPT \cite{wachter2006implementation} and the convergence tolerance is set to $0.0001$. We assume IPOPT returns a feasible solution, which may not be a global optimum.
We test our algorithm on different sizes of IEEE networks. For the 9-bus and 22-bus networks, we utilize the local solutions found in \cite{bukhsh2013local} as starting points to launch IPOPT. Specifically, there are 4 local solutions in the 9-bus network, and 2 local solutions in the 22-bus network. For each network, starting from different local solutions, our algorithm can achieve global solutions within one iteration.

\begin{figure}[t]
\centering
\includegraphics[scale=0.4]{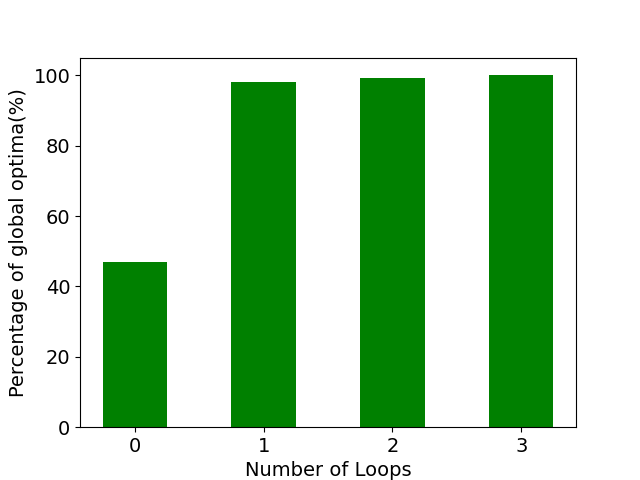}
\caption{Using a set of random starting points, 47\% of them leads to the global optimal after a direct call to IPOPT. The fraction of global optimal solutions increases to 98\%, 99.93\% and 100\% after one, two and three iterations of algorithm 1, respectively. 
\vspace{-0.5cm}}
\label{fig:39bus}
\end{figure}

For the 39-bus network, we take a set of $600$ initial points for $\btheta$ and $\bV$. To find the global solution, we do an exhaustive search within the bounds of each variable.
In Fig. \ref{fig:39bus}, we illustrate the improvement of solution quality for the 39-bus network. The x-axis represents the number of iterations that Algorithm 1 is ran, and y-axis represents the percentage of globally optimal solutions after each iteration. Without Algorithm 1, less than half of the solutions are globally optimal. One application of Algorithm 1 increases the percentage of globally optimal solutions to $98\%$.
After two iterations, only four cases do not reach global optimas. 
All solutions are globally optimal after three iterations. 

\captionsetup[figure]{font=small,skip=2pt}

\section{Conclusion}
\label{sec:conclusion}
In this paper, we propose a simple algorithm to find globally optimal solutions to ACOPF problems iteratively.
First, we solve the ACOPF problem using an existing nonlinear solver. From the solution and its associated dual variables, we construct a partial Lagrangian. Optimizing this partial Lagrangian leads to a new solution. With this solution as an initial point, we again call the solver for the ACOPF problem. By repeating these steps, we can iteratively improve the solution quality, escaping from local solutions to achieve global optimums. We illustrate the intuition behind our algorithm using a two-bus network, which shows that the partial Lagrangian has a flatter optimization landscape compared to the original primal problem. 
We validate the effectiveness of our algorithm on standard 9-bus, 22-bus and 39-bus networks. 
Regardless of the initial points, our algorithm always finds the global optimum within at most three iterations.

\bibliographystyle{IEEEtran}
\bibliography{mybib.bib}

\end{document}